\documentclass[%
 reprint,
 amsmath,amssymb,
 aps,
]{revtex4-1}

\usepackage{graphicx}
\usepackage{dcolumn}
\usepackage{bm}

\begin{document}


\title{Effect of Sterile Neutrinos and Nonstandard Interactions on the Geo-neutrino Flux}

\author{Gwanwen Yan}
\affiliation{Dept of physics, University of Maryland, College Park, Md-20742
}%

\date{\today}

\begin{abstract} We calculate the effect of sterile neutrinos and nonstandard neutrino interactions on the flux of active neutrinos from the Earth's crust and mantle taking matter effect into account in both cases. For simple Earth model and using previous emission flux estimates at the source, we find that for reasonable choice of parameters for both the sterile neutrino and nonstandard interaction, the effect on the neutrino flux is a few percent or less.
\end{abstract}

\pacs{Valid PACS appear here}
\maketitle


\section{Introduction}
There is considerable discussion in the literature about the possibility that there exist very light sterile neutrinos (with mass less than or of the order of an eV) that mix with the known three active neutrinos \cite{review} \cite{sterile3}\cite{sterile5}. While none of the indications for such neutrinos is conclusive, the existence of such particle has such significant impact on the nature of physics beyond the standard model that all possible venues where their effects may manifest must be explored.\cite{sterile4}  The existence of sterile neutrinos affects the propagation of active neutrinos in all  experimental settings.\cite{sterile2}\cite{icecube}\cite{indirect} \\
One of the ways to verify the existence of sterile neutrinos is to analyze geoneutrino fluxes. Geoneutrinos are neutrinos being produced by radioisotopes inside the Earth.\cite{mantle}\cite{mcdonough} There are several geoneutrino detectors around the world, such as jinping underground laboratory \cite{jinping2} and KamLAND \cite{kamland2}. There is another neutrino detector under construction in Jiangmen, China\cite{juno}. By calculating the geoneutrino flux on the surface of the Earth using various neutrino oscillation models, we are able to find out which oscillation model matches the experimental data the best.

\section{Origion of Geoneutrinos}
There are three main sources of neutrino that can be detected on Earth: cosmological neutrinos\cite{cosmo}, reactor neutrinos\cite{reactor} \cite{reactor2}, and geoneutrinos\cite{geo2}. In this paper we are focusing on geoneutrinos, which come from inside the Earth. There are several kinds of radioisotopes in the earth that can produce neutrino fluxes. Three major contributions of geo-neutrino fluxes are $^{238}U$,$^{232}Th$, and $^{40}K$. The decay chain of these three isotopes are\cite{jinping}:
\begin{equation}
\begin{array}{llll}
^{238}U\to ^{206}Pb+8\alpha+6e^-+6\bar{\nu}_e+51.698MeV\\
^{232}Th\to ^{207}Pb+7\alpha+4e^-+4\bar{\nu}_e+46.402MeV\\
^{40}K\to ^{40}Ca+e^-+4\bar{\nu}_e+1.311MeV (89.3 \%) \\
^{40}K+e^-\to ^{40}Ar+\nu_e+1.505MeV (10.7 \%)\\
\end{array}
\end{equation}
However, the energy produced by the two potassium decay chains are below the IBD (inverse beta decay) threshold of 1.8 MeV. So the neutrino flux being produced by those processes cannot be detected by current detectors.\cite{jinping}\\
 
\section{The Earth Model}
The earth model we used is CRUST1.0\cite{crust} In that model, the Earth contains seven layers: upper crust, lower crust, upper mantle, transition zone, lower antle, outer core, and inner core\cite{kamland}. The density and the abundance of HPEs are assumed to be uniform in each layer. \cite{earthmodel}The depth, density and HPE abundances in each layer are listed in Table 1.\cite{earth}

\begin{table}[]
\centering
\caption{Earth Model}
\label{Table 1}
\begin{tabular}{lllll}
\hline
Layer                & Depth & Density & Abundance(U) & Abundance(Th) \\
                     & $km$    & $g/cm^3$   & $ppm$          & $ppm$           \\
\hline
Surface of the Earth & 0     & 0       & 0            & 0             \\
Upper Crust          & 15    & 2.6     & 2.8          & 10.7          \\
Lower Crust          & 25    & 2.9     & 0.2          & 1.2           \\
Upper Mantle         & 220   & 3.36    & 0.012        & 0.048         \\
Transition Zone      & 400   & 3.54    & 0.012        & 0.048         \\
Lower Mantle         & 670   & 3.99    & 0.012        & 0.048         \\
Outer Core           & 2891  & 5.57    & 0            & 0             \\
Inner Core           & 5150  & 12.17   & 0            & 0             \\
\hline            
\end{tabular}
\end{table}

\section{Neutrino Oscillation in vacuum}
We used PMNS matrix to describe the relation between mass and flavor eigenstates of neutrinos. Consider a neutrino with a specific lepton flavor $|\nu_a\rangle$, where $a=e, \mu, \tau$. In general, this flavor eigenstate of a neutrino is not a mass eigenstate, but rather a superposition of mass eigenstates $|\nu_i \rangle$ , $i=1,2,3$

\begin{equation}
|\nu_a \rangle =\sum_i U_{ai} |\nu_i \rangle
\end{equation}

For the active neutrino model, we have
$$
\begin{aligned}
U&=
\begin{pmatrix}
 U_{e1} & U_{e2} & U_{e3} \\
 U_{\mu 1} & U_{\mu 2} & U_{\mu 3} \\
 U_{\tau 1} & U_{\tau 2} & U_{\tau 3} \\
\end{pmatrix}\\
&=\begin{pmatrix}
 1 & 0 & 0 \\
 0 & c_{23} & s_{23} \\
 0 & -s_{23} & c_{23} \\
\end{pmatrix}
\begin{pmatrix}
 c_{13} & 0 & s_{13} \\
 0 & 1 & 0 \\
 -s_{13} & 0 & c_{13} \\
\end{pmatrix}
\begin{pmatrix}
 c_{12} & s_{12} & 0 \\
 -s_{12} & c_{12} & 0 \\
 0 & 0 & 1 \\
\end{pmatrix} \\
&=\begin{pmatrix}
 c_{12}c_{13} & -s_{12}c_{13} & s_{13} \\
 s_{12}c_{23}+c_{12}s_{23}s_{13} & c_{12}c_{23}-s_{12}s_{23}s_{13} & -s_{23}c_{13} \\
 s_{12}s_{23}-c_{12}c_{23}s_{13} & c_{12}s_{23}+s_{12}c_{23}s_{13} & c_{23}c_{13} \\
\end{pmatrix}
\end{aligned}
$$

Here $s_{ij}=\sin\theta_{ij}$, $c_{ij}=\cos\theta_{ij}$. $\theta_{ij}$ is the mixing angle between mass eigenstates $i$ and $j$. \\
If we set $\theta_{12}=0.584$,$\theta_{13}=0.149$, and $\theta_{23}=0.785$, the relation between mass and lepton eigen states becomes

\begin{equation}
U_3=
\left(
\begin{array}{cccc}
 0.825 & 0.545 & 0.148 & 0 \\
 -0.478 & 0.532 & 0.699 & 0 \\
 0.302 & -0.648 & 0.670 & 0 \\
 0 & 0 & 0 & 1 \\
\end{array}
\right)
\end{equation}

Similarly, if we add the sterile neutrino into our model, The analytic form of U becomes:

\[\begin{aligned}
 U_{e1}&=c_{12} c_{13} c_{14} \\
 U_{e2}&=c_{13} c_{14} s_{12} \\
 U_{e3}&=c_{14} s_{13} \\
 U_{e4}&=s_{14} \\
 U_{\mu 1}&=-c_{23} c_{24} s_{12} + c_{12} (-c_{24} s_{13} s_{23} - c_{13} s_{14} s_{24}) \\
 U_{\mu 2}&= c_{12} c_{23} c_{24} + s_{12} (-c_{24} s_{13} s_{23} - c_{13} s_{14} s_{24}) \\
 U_{\mu 3}&= c_{13} c_{24} s_{23} - s_{13} s_{14} s_{24} \\
 U_{\mu 4}&= c_{14} s_{24} \\
 U_{\tau 1}&= -s_{12} (-c_{34} s_{23} - c_{23} s_{24} s_{34}) \\
           &+ c_{12} [-c_{13} c_{24} s_{14} s_{34} - s_{13} (c_{23} c_{34} - s_{23} s_{24} s_{34})]\\
 U_{\tau 2}&= c_{12} (-c_{34} s_{23} - c_{23} s_{24} s_{34}) \\
           &+ s_{12} [-c_{13} c_{24} s_{14} s_{34} - s_{13} (c_{23} c_{34} - s_{23} s_{24} s_{34})] 
             \end{aligned}\]
 \[\begin{aligned}
 U_{\tau 3}&= -c_{24} s_{13} s_{14} s_{34} + c_{13} (c_{23} c_{34} - s_{23} s_{24} s_{34}) \\
 U_{\tau 4}&= c_{14} c_{24} s_{34} \\
 U_{s1}&= -s_{12} (-c_{23} c_{34} s_{24} + s_{23} s_{34}) \\
       &+ c_{12} [-c_{13} c_{24} c_{34} s_{14} - s_{13} (-c_{34} s_{23} s_{24} - c_{23} s_{34})] \\
 U_{s2}&= c_{12} (-c_{23} c_{34} s_{24} + s_{23} s_{34}) \\
       & + s_{12} [-c_{13} c_{24} c_{34} s_{14} - s_{13} (-c_{34} s_{23} s_{24} - c_{23} s_{34})] \\
 U_{s3}&= -c_{24} c_{34} s_{13} s_{14} + c_{13} (-c_{34} s_{23} s_{24} - c_{23} s_{34}) \\
 U_{s4}&= c_{14} c_{24} c_{34} \\
 \end{aligned}\]
 
In this mode, there are 6 mixing angles: $\theta_{12}$, $\theta_{13}$, $\theta_{14}$, $\theta_{23}$, $\theta_{24}$, $\theta_{34}$.
Here we set $\theta_{12}=0.584$, $\theta_{13}=0.149$, $\theta_{23}=0.785$, $\theta_{14}=\theta_{24}=\theta_{34}=0.1$ \\ Then the relation between mass and lepton eigenstates becomes: \\

\begin{equation}\label{uvac4}
\left(
\begin{array}{cccc}
\nu_e \\
\nu_{\mu} \\
\nu_{\tau} \\
\nu_s \\
\end{array}
\right)
=
\left(
\begin{array}{cccc}
 0.821 & 0.542 & 0.148 & 0.100 \\
 -0.483 & 0.524 & 0.694 & 0.099 \\
 0.297 & -0.655 & 0.688 & 0.099 \\
 -0.064 & -0.042 & -0.154 & 0.985 \\
\end{array}
\right)
\left(
\begin{array}{cccc}
\nu_1 \\
\nu_2 \\
\nu_3 \\
\nu_4 \\
\end{array}
\right)
\end{equation}

From equation \ref{uvac4} and the definition of U, we can derive the transition probability between two neutrino flavors:
\begin{equation}
P_{\alpha\beta}(L)=\left|\sum_{i=i}^3 U_{\alpha i}^* U_{\beta i}e^{-i\frac{m_i^2L}{2E}}\right|^2
\end{equation}
Then the electron survival probability becomes
\begin{equation}\label{eqn:pee}
P_{ee}(E,L)=1-4\sum_{i>j}\left| U_{ei}\right|^2 \left| U_{ej}\right|^2 sin^2(\frac{\Delta m_{ij}^2L}{4E})
\end{equation}

In general, there might be CP phases in neutrino mixing and the matrix elements $U_{\alpha i}$ are complex. Those CP phases might affect the transition probabilities of different flavors of neutrinos. However, in this paper we focus only on the survival probability of electron neutrinos. The relevant matrix elements are $U_{ei}$, $i=1,2,3,4$. All these elements are multiplication of trigonometric functions, and the CP phases do not affect the mode squares of such elements.

\section{Geoneutrino Oscillation inside the Earth}
Neutrino oscillation can be described by the following differential equation:

\begin{equation}\label{hamiltonian}
i\frac{d}{dt}
\left(
\begin{matrix}
 \nu_e \\
 \nu_{\mu} \\
 \nu_{\tau} \\
 \nu_s \\
\end{matrix}
\right)
= \hat{H}
\left(
\begin{matrix}
 \nu_e \\
 \nu_{\mu} \\
 \nu_{\tau} \\
 \nu_s \\
\end{matrix}
\right)
\end{equation}

When neutrinos travel through matter, there will be a potential term in the hamiltonian $\hat{H}$ that will affect the oscillation. There are two contributions to the potential. One is called Mikheyev-Smirnov-Wolfenstein effect, or MSW effect.\cite{msw} This part of potential comes from the scattering between electrons and neutrinos. The contribution of MSW effect to the Hamiltonian is:

\begin{equation}
\begin{aligned}
V_{MSW}=&\sqrt{2}G_f n_e
\begin{bmatrix}
 1 & 0 & 0 & 0 \\
 0 & 0 & 0 & 0 \\
 0 & 0 & 0 & 0 \\
 0 & 0 & 0 & 0 \\
\end{bmatrix}\\
&+\frac{\sqrt{2}}{2}G_f n_n
\begin{bmatrix}
 1 & 0 & 0 & 0 \\
 0 & 1 & 0 & 0 \\
 0 & 0 & 1 & 0 \\
 0 & 0 & 0 & 0 \\
\end{bmatrix}
\end{aligned}
\end{equation}

The other contribution is from non-standard interactions (NSI)\cite{amir}\cite{liao}, which has the form in the effective potential:\cite{nsi} 

\begin{equation}
V_{NSI}=\sqrt{2}G_f n_e
\begin{bmatrix}
 \epsilon_{ee} & \epsilon_{e\mu} & \epsilon_{e\tau} & 0 \\
 \epsilon_{\mu e} & \epsilon_{\mu\mu} & \epsilon_{\mu\tau} & 0 \\
 \epsilon_{\tau e} & \epsilon_{\tau\mu} & \epsilon_{\tau\tau} & 0 \\
 0 & 0 & 0 & 0 \\
\end{bmatrix}
\end{equation}

After adding these two potentials to $\hat{H}$, equation \ref{hamiltonian} becomes

\begin{equation}
\begin{aligned}
i\frac{d}{dt}
\begin{bmatrix}
 \nu_e \\
 \nu_{\mu} \\
 \nu_{\tau} \\
 \nu_s \\
\end{bmatrix}
=& \frac{1}{2E}\left( U
\begin{bmatrix}
 0 & 0 & 0 & 0 \\
 0 & \Delta m_{21}^2 & 0 & 0 \\
 0 & 0 & \Delta m_{31}^2 & 0 \\
 0 & 0 & 0 & \Delta m_{41}^2 \\
\end{bmatrix}
U^{\dagger}\right.\\
&\left.+2\sqrt{2}EG_f n_e
\begin{bmatrix}
 1+\epsilon_{ee} & \epsilon_{e\mu} & \epsilon_{e\tau} & 0 \\
 \epsilon_{\mu e} & \epsilon_{\mu\mu} & \epsilon_{\mu\tau} & 0 \\
 \epsilon_{\tau e} & \epsilon_{\tau\mu} & \epsilon_{\tau\tau} & 0 \\
 0 & 0 & 0 & 0 \\
\end{bmatrix}\right.\\
&\left.
-\sqrt{2}EG_f n_n
\begin{bmatrix}
 1 & 0 & 0 & 0 \\
 0 & 1 & 0 & 0 \\
 0 & 0 & 1 & 0 \\
 0 & 0 & 0 & 0 \\
\end{bmatrix}
\right)
\begin{bmatrix}
 \nu_e \\
 \nu_{\mu} \\
 \nu_{\tau} \\
 \nu_s \\
\end{bmatrix}
\end{aligned}
\end{equation}

Diagonalizing the above matrix, we have

\begin{equation}
\hat{H}=\frac{1}{2E}U_m\text{diag}(m_1^2,m_2^2,m_3^3,m_4^2)U_m^{\dagger}
\end{equation}

After plugging in the upper limit of NSIs for geoneutrinos:\cite{nsi}

\begin{equation} \label{nsis}
\left(
\begin{array}{cccc}
 \vert\epsilon_{ee}\vert<4.2 & \vert\epsilon_{e\mu}\vert<0.33 & \vert\epsilon_{e\tau}\vert<3.0 & \vert\epsilon_{es}\vert=0 \\
  & \vert\epsilon_{\mu\mu}\vert<0.068 & \vert\epsilon_{\mu\tau}\vert<0.33 & \vert\epsilon_{\mu s}\vert=0 \\
  &  & \vert\epsilon_{\tau\tau}\vert<21 & \vert\epsilon_{\tau s}\vert=0 \\
  &  &  & \vert\epsilon_{ss}\vert=0 \\
\end{array}
\right)
\end{equation}

The transition matrix becomes
$$
U_m=
\left(
\begin{array}{cccc}
 0.820 & 0.544 & 0.148 & 0.100 \\
 -0.484 & 0.524 & 0.694 & 0.099 \\
 0.298 & -0.655 & 0.688 & 0.099 \\
 -0.064 & -0.042 & -0.154 & 0.985 \\
\end{array}
\right)
$$

\section{Calculation of Geoneutrino Flux}
In this section we calculated the total geoneutrino flux that should be detected on Earth. Integrating the $\bar{\nu}_e$ flux produced from all seven layers inside the Earth, we get the expression of the $\bar{\nu}_e$ flux for each element: \cite{jinping}
\begin{equation}
\phi_i=\frac{\lambda_i N_A}{\mu_i}n_{\nu}(i)\int P_{ee}(L) \frac{A(\vec{r})\rho(\vec{r})}{4\pi L^2}d\vec{r}
\end{equation}

Here $\lambda$ is the decay constant, $\mu$ is the standard atomic molar mass, $n_{\nu}$ is the number of $\vec{\nu}_e$ emitted per decay, A is the natural abundance of the element, L is the distance between the source and the detector, i can be U, Th, or K. \\
Here we assumed uniform abundance of HPE's in mantle, and no HPE's in the core of the Earth.\cite{mantle}That is probably the reason why the total flux we calculated is smaller than observed values. We used a seven-layer model of the earth, and assumed uniform density in each layer.\cite{earth}
The average electron neutrino survival probability for a specific neutrino energy E is \cite{msw}

\begin{equation}
P_{ee}(E)=\frac{\int P_{ee}(E,L)A(\vec{r})\rho(\vec{r})/(4\pi L^2)d\vec{r}}{\int A(\vec{r})\rho(\vec{r})/(4\pi L^2)d\vec{r}}
\end{equation}

Where $P_{ee}(E,L)$ can be calculated from equation (\ref{eqn:pee}) \\

\begin{figure}
\centering
\includegraphics[width=0.4\textwidth]{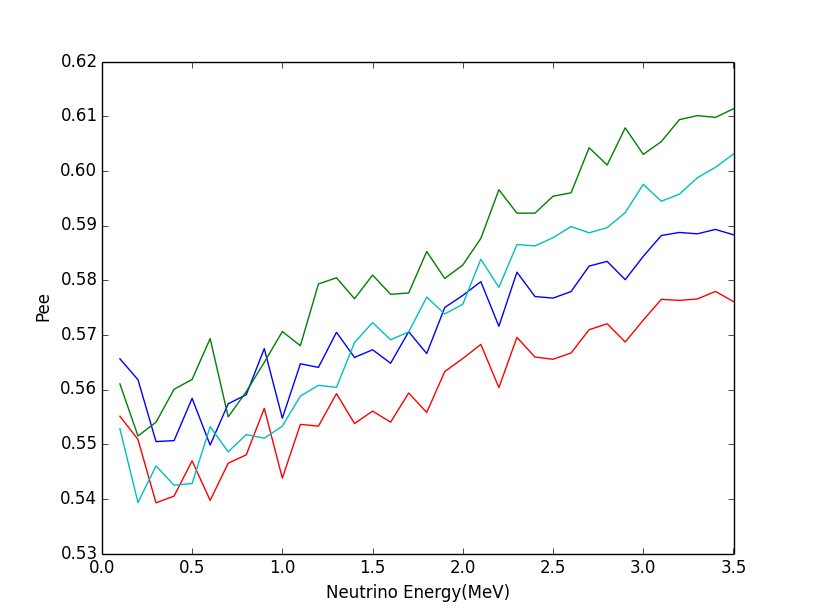}
\caption{Pee vs Neutrino energy. Blue: 3-neutrino model in vacuum; Green: 3-neutrino model in matter with NSI; Red: 4-neutrino model in vacuum; Pink: 4-neutrino model in matter with NSI}
\label{fig:Pee}
\end{figure}

\begin{figure}
\centering
\includegraphics[width=0.4\textwidth]{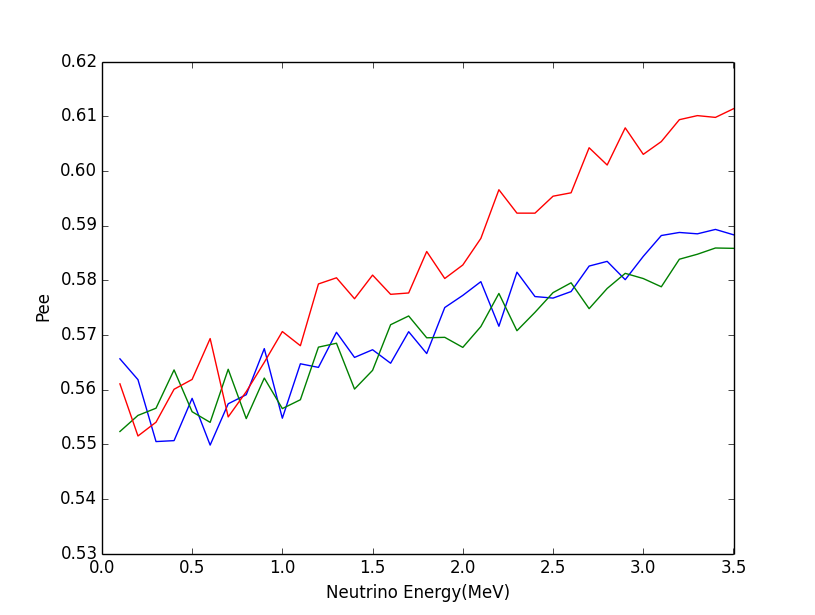}
\caption{Pee vs Neutrino energy for 3-neutrino model. Blue: in vacuum; Green: in matter without NSI; Red: in matter with NSI}
\label{fig:Pee3}
\end{figure}

\begin{figure}
\centering
\includegraphics[width=0.4\textwidth]{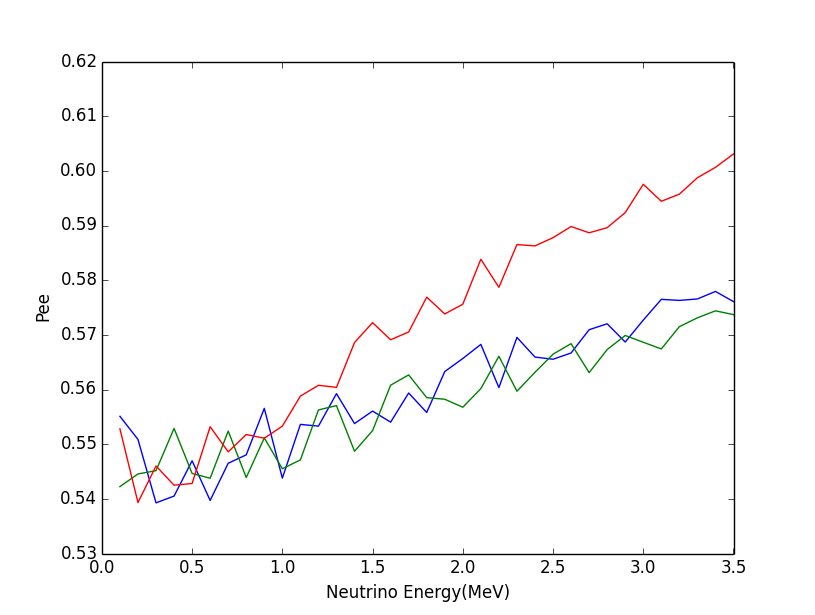}
\caption{Pee vs Neutrino energy for 4-neutrino model. Blue: in vacuum; Green: in matter without NSI; Red: in matter with NSI}
\label{fig:Pee4}
\end{figure}

From the plots, we can see that the electron survival probabilities increase when energy increases, and the MSW effect has very small effect on geoneutrinos. These agrees with previous results.\cite{jinping} Furthermore, we can see that if we add a sterile neutrino to the 3-neutrino model, $P_{ee}$ will decrease by about 2\% . But if we also consider the non-standard interactions, $P_{ee}$ will increase by about 2\% .

Gianni Fiorentini etc. provided an analytic way to calculate geoneutrino flux.\cite{geo} In their model, the total geoneutrino flux $\phi(X)$ can be estimated as:

\begin{equation}
\begin{aligned}
\phi(X)&=\frac{A_XR_e}{2}\left[ \frac{R_2}{R_e}-\frac{1}{2}\frac{R_e^2-R_2^2}{R_e^2}\log\left(\frac{R_e+R_2}{R_e-R_2}\right)\right.\\&\left.-\frac{R_1}{R_e}+\frac{1}{2}\frac{R_e^2-R_1^2}{R_e^2}\log\left(\frac{R_e+R_1}{R_e-R_1}\right)\right]
\end{aligned}
\end{equation}

Where A is the specific geo-neutrino activity, i.e. the number of geo-neutrinos produced per unit time and volume, X stands for U or Th, $R_e$ is the radius of the earth, $R_1$ and $R_2$ are the radius of lower and upper bounds of a spherical layer of the Earth.
Using this equation, the total geoneutrino flux in our earth model is 44.4TNU. This value is very close to the numerical result from computer simulation (45TNU).

\section{Conclusion}

\begin{figure}
\centering
\includegraphics[width=0.4\textwidth]{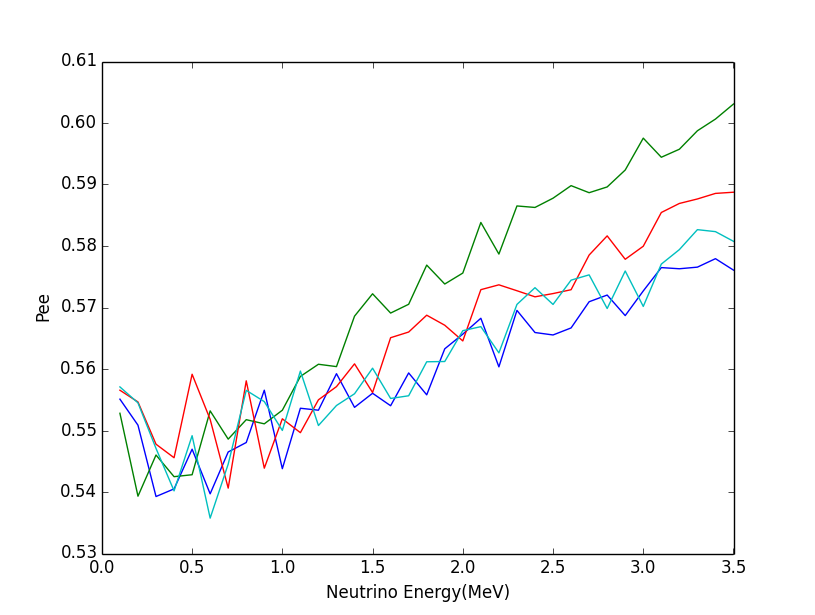}
\caption{Pee vs Neutrino energy for 4-neutrino model with different strength of NSI. Blue: without NSI; Green: the upper limit of NSI; Red: 1/2 the upper limit of NSI; Purple: 1/4 of the upper limit of NSI}
\label{fig:nsi}
\end{figure}

We analyzed the oscillation of electron antineutrinos inside the Earth using the simplified Earth model and taking account both the matter effect and the NSI effect. The results are shown in three figures.
In Fig.\ref{fig:Pee} we can see that the average electron antineutrino survival probabilities($P_{ee}$) under different conditions are between 0.54 and 0.61. They increase as neutrino energies increase. This coincides with the prediction in Wan's Paper.\cite{jinping} Comparing the blue line with the red line, we can see that if we include sterile neutrinos into the oscillation model, $P_{ee}$ will become lower than that of the three-neutrino model. This phenomenon is expected, since the extra type of neutrino produces an extra degree of freedom in the flavor mixing. \\
However, if we include the non-standard interactions(NSI), $P_{ee}$ becomes larger. Fig.\ref{fig:Pee3} and Fig.\ref{fig:Pee4} show the effect of Earth matter and NSI to $P_{ee}$. Comparing the blue lines and the green lines in these two figures, we can see that the MSW effect of the Earth is very small. The NSI effect, on the other hand, is larger than MSW effect, and increases $P_{ee}$. The change in $P_{ee}$ caused by NSI becomes larger as the energy of neutrino increases. The average change is about 2\% with the strongest possible effect of NSI. \cite{potential} \cite{bounds} \cite{darkmatter} \\
In the previous discussion, we plugged in the highest possible values allowed by experimental observation in equation \ref{nsis}. In general, the actual matrix elements can be complex and the effect on $P_{ee}$ might be much smaller then what is showed in figure \ref{fig:Pee3} and figure \ref{fig:Pee4}. We reduced the values of all elements in equation \ref{nsis} by a factor of 2 and 4, and examined the change of $P_{ee}$. The results are showed in figure \ref{fig:nsi}. We can see that the effect of NSI on $P_{ee}$ is roughly proportional to the values of matrix elements. When the values are below 1/4 of the experimental bounds, the effect of NSI will be hard to see.

\section{acknowledgement}
I wish to thank Prof. Rabindra Mohapatra for suggesting the problem, valuable discussions and kind suggestions on this paper.

\bibliographystyle{plain}

\end{document}